\documentclass[
    pra,
    aps,
    reprint,
    amsmath,amssymb,
    a4paper,
    superscriptaddress,
    10pt
]{revtex4-2}
\usepackage[english]{babel}
\usepackage{graphicx}
\usepackage{dcolumn}
\usepackage{bm}
\usepackage{textcomp}
\usepackage{upgreek}
\usepackage{amsfonts}
\usepackage{mhchem}
\usepackage{physics}
\usepackage{gensymb}
\usepackage[dvipsnames,table,xcdraw]{xcolor}

\definecolor{rmpblue}{HTML}{2e3092}

\usepackage[
    colorlinks=true,
    citecolor=rmpblue,
    linkcolor=rmpblue,
    filecolor=magenta,
    urlcolor=rmpblue
]{hyperref}

\graphicspath{{Fig/}}

\newcommand{\affilANU}{Research School of Physics, Australian National University, Canberra ACT 2601, Australia}

\newcommand{\affilUTS}{School of Mathematical and Physical Sciences, University of Technology Sydney, Ultimo, New South Wales 2007, Australia}

\newcommand{\affilPostechChe}{Department of Chemical Engineering, Pohang University of Science and Technology (POSTECH), Pohang, Republic of Korea}

\newcommand{\affilPostechMec}{Department of Mechanical Engineering, Pohang University of Science and Technology (POSTECH), Pohang 37673, Republic of Korea}

\newcommand{\affilPostechEle}{Department of Electrical Engineering, Pohang University of Science and Technology (POSTECH), Pohang 37673, Republic of Korea}

\newcommand{\affilPostechQua}{Quantum Information Science Program, Graduate School of Convergence Science and Technology, Pohang University of Science and Technology (POSTECH), Pohang 37673, Republic of Korea}

\newcommand{\affilPostechIns}{National Institute of Nanomaterials Technology (NINT), Pohang 37673, Republic of Korea}

\begin{document}

\title[]{Metaphotonic structures with strong linear and nonlinear chiral response}

\author{Pavel Tonkaev}
\email{pavel.tonkaev@anu.edu.au}
\thanks{These authors contributed equally.}
\affiliation{\affilANU}
\affiliation{\affilUTS}

\author{Jihae Lee}
\thanks{These authors contributed equally.}
\affiliation{\affilPostechChe}

\author{Ivan Toftul}
\affiliation{\affilANU}

\author{Brijesh Kumar}
\affiliation{\affilANU}

\author{Junsuk Rho}
\affiliation{\affilPostechChe}
\affiliation{\affilPostechMec}
\affiliation{\affilPostechEle}
\affiliation{\affilPostechQua}
\affiliation{\affilPostechIns}

\author{Yuri Kivshar}
\email{yuri.kivshar@anu.edu.au}
\affiliation{\affilANU}

\begin{abstract}
The symmetry of a structure plays a fundamental role in determining its optical response, particularly its interaction with circularly polarized light, with implications for applications ranging from polarization control to enantioselective sensing. Dielectric metasurfaces can enhance chiroptical effects, but most approaches rely on geometrically chiral building blocks that are difficult to design and fabricate. In addition, chiral effects in nonlinear optical structures remain relatively unexplored. Here we show that a suspended, substrate-free membrane metasurface with $C_4$-symmetric apertures, which is only two-dimensionally chiral, becomes three-dimensionally chiral when a thin silicon layer (remaining after incomplete etching) breaks its out-of-plane mirror symmetry. The layer enables hybridization of resonant membrane modes of opposite parity, yielding a large linear circular dichroism observed in experiment. Nonlinear circular dichroism requires only two-dimensional chirality of the structure including its crystal lattice. In the partially etched membrane, the nonlinear response is found to be considerably stronger, with third-harmonic generation under right-circularly polarized excitation enhanced by more than two orders of magnitude compared with an unpatterned film and exceeding that under left-circularly polarized excitation by more than an order of magnitude,  corresponding to a nonlinear circular dichroism of up to 0.91. Our results suggest that suspended partially etched membranes can provide a minimal platform for chiral nonlinear photonics, with potential for handedness-selective frequency conversion and compact chiral light sources.
\end{abstract}

\maketitle

The interaction of light with chiral structures provides a mean of controlling optical responses according to the handedness of the incident electromagnetic field. When illuminated by left- and right-circularly polarized (LCP and RCP) light, three-dimensionally chiral media can exhibit different transmission, reflection, and absorption characteristics, giving rise to characteristic chiroptical effects such as circular dichroism (CD) and optical rotatory dispersion~\cite{berova2000circularDichroism,Barron_optical_activity_2004}. These polarization-dependent responses are of considerable interest for applications in enantioselective sensing~\cite{Solomon2019:ACSPhot:dielectricchiral}, polarization optics~\cite{Arbabi2015,BalthasarMueller2017}, spin-selective light–matter interactions~\cite{Mak_chiral_int_2012,Yang_Chiral_light_matter_int_2019}, and nonlinear chiral photonics. Achieving strong and controllable optical chirality at the subwavelength scales is therefore an important objective in the development of compact photonic structures and devices for chiral sensing~\cite{yoo_metamaterials_2019, chen_integrated_2020}, chiral lasing~\cite{lin_toward_2023, deng_chiral_2025}, and selective chiral transmission~\cite{yang2019spin, kuhner2023unlocking,gryb2023two, duan_spin-preserving_2024,sinev2025chirality}. 

Natural materials generally exhibit relatively weak chiroptical responses, motivating extensive efforts to enhance optical chirality using artificially structured materials~\cite{pendry2004a,wang2009chiral,oh2015chiral}. In particular, metasurfaces have emerged as a versatile platform for controlling light–matter interactions through the engineering of resonant electromagnetic modes, local field distributions, and structural symmetries within deeply subwavelength unit cells. All-dielectric metasurfaces are particularly attractive because they support low-loss resonances and strong electromagnetic field confinement~\cite{Neshev2018,Koshelev_meta2019,Kumar2023,Hu2025}. By exploiting resonant modes, including chiral quasi-bound states in the continuum (qBICs), these structures can achieve pronounced CD and approach the regime of maximum optical chirality~\cite{Gorkunov2020PRL,shi_planar_2022}, in which one circular polarization is strongly suppressed while the opposite handedness is efficiently transmitted or reflected. Nevertheless, many existing approaches rely on geometrically chiral meta-atoms, which often require complex three-dimensional geometries and introduce additional challenges in design and fabrication~\cite{Gansel2009,Tseng2019}.

\begin{figure*}[!htbp]
    \centering
    \includegraphics[width=0.99\linewidth]{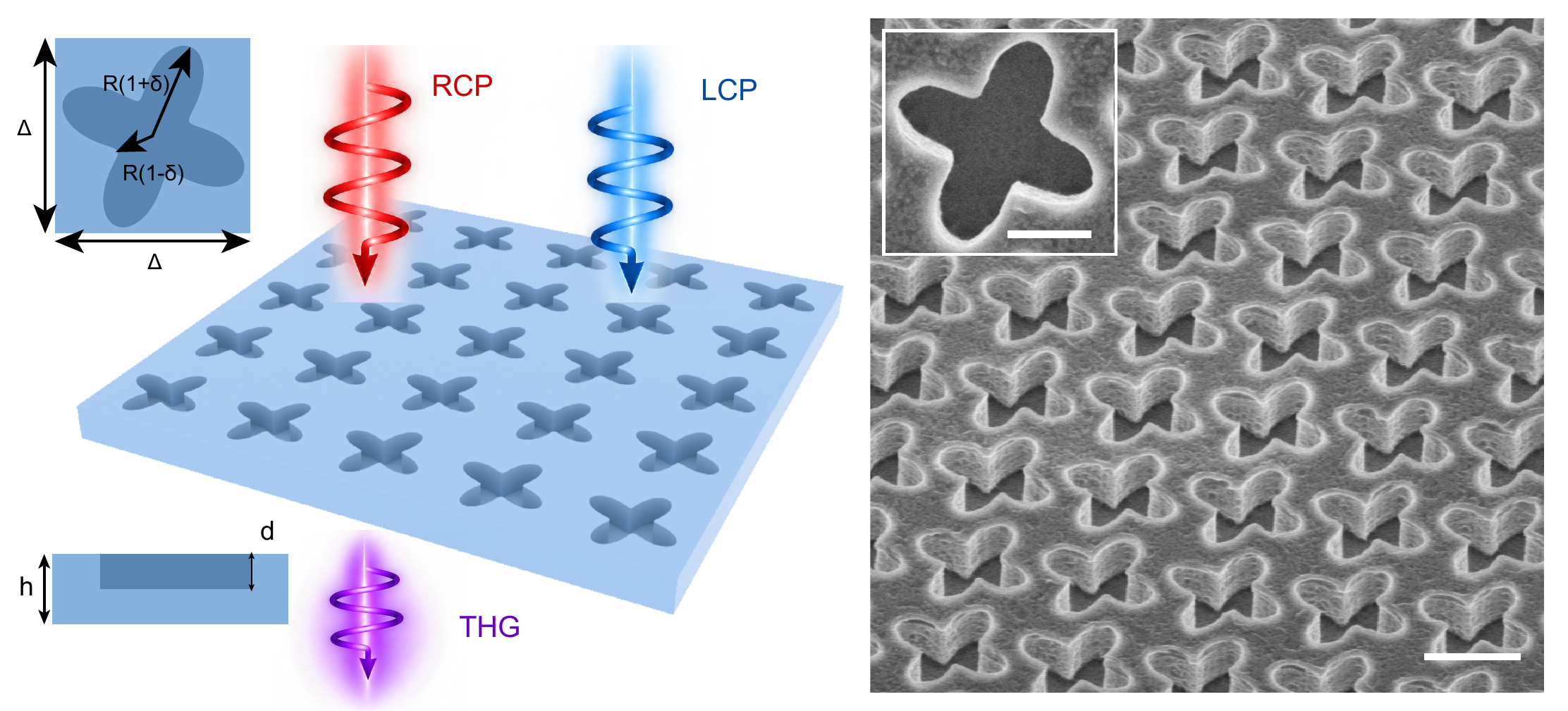}
    \caption{\textbf{Concept of chiral metasurface.} (a) Schematic of the membrane metasurface and its unit cell (top view and cross-section). The unit cell has period $\Delta$ and consists of four-lobed (C$_4$) etched features built from ellipses with radii $R(1\pm\delta)$, where $\delta$ is the asymmetry parameter. The etch depth is $d$ and the membrane thickness is $h$. RCP excitation produces strong THG, whereas LCP excitation produces weak THG. (b) Scanning electron microscopy image of the fabricated metasurface. Scale bar: 500~nm.}
    \label{fig:concept}
\end{figure*}

Beyond complex three-dimensional meta-atoms, breaking the out-of-plane mirror symmetry of a two-dimensionally chiral pattern provides an alternative route to three-dimensional chirality and to controlling the optical response of metasurfaces.
Such symmetry breaking is relevant not only to chiral photonics but also to a broader range of topology-driven phenomena, including unidirectional guided resonances, asymmetric radiation in Janus metasurfaces, and the engineering of bound states in the continuum and exceptional points in non-Hermitian photonic systems~\cite{yin2023topological,kang2025janus,su2026continuous}. In particular, previous theoretical studies have demonstrated that introducing a second layer beneath a membrane metasurface reduces its symmetry from C$_{4h}$ to C$_4$, enabling the hybridisation of resonant modes with opposite parity. Combined with appropriately engineered dissipation, this mechanism can give rise to maximal optical chirality~\cite{Kumar2025chiral}. This concept was subsequently demonstrated experimentally using free-standing silicon membrane metasurfaces, where a spin-coated poly(methyl methacrylate) (PMMA) layer breaks the out-of-plane mirror symmetry and produces a substantial linear CD~\cite{Kumar2026}. However, this approach relies on an additional material layer, whose thickness and associated optical losses can be difficult to control precisely. An alternative strategy that achieves controlled symmetry breaking directly through the membrane geometry, without introducing an additional material, would therefore be highly desirable.

In parallel with advances in linear chiral photonics, {\it nonlinear chiroptical effects} have attracted increasing interest as a means of enhancing and controlling light–matter interactions~\cite{koshelev2023nonlinear}. In particular, resonantly enhanced nonlinear processes can produce pronounced CD and helicity-selective harmonic generation, offering functionalities beyond those accessible in the linear regime~\cite{tonkaev_nonlinear_2024,toftul_chiral_2024,toftul_monoclinic_2025,tonkaev_nonlinear_2025,lai2025nonlinear,heimig2026chiral}. Recent studies have further demonstrated that strong nonlinear circular dichroism can emerge even in planar membrane metasurfaces that preserve out-of-plane mirror symmetry, where two-dimensional chirality is sufficient (no in-plane mirror line maps the structure, including its crystal lattice, onto itself), through the interplay between lattice symmetry, meta-atom orientation, crystal orientation, and resonant optical modes~\cite{li2026meta,hariharan2026nonlinear,Wang2026NanoLett}. These developments highlight the potential of metasurfaces for engineering nonlinear chiroptical responses without relying on intrinsically three-dimensional chiral structures. However, the role of out-of-plane symmetry breaking in controlling resonant mode hybridisation and especially nonlinear chiral frequency conversion remains largely unexplored.

Here, we demonstrate experimentally that a suspended, substrate-free silicon membrane metasurface, which exhibits no linear circular dichroism when its out-of-plane mirror symmetry is preserved, can be transformed into {\it a strongly chiral optical platform} through partial etching of its apertures. The residual silicon layer beneath the patterned region breaks the out-of-plane mirror symmetry reducing the point-group symmetry from (C$_{4h}$) to (C$_4$), enabling the hybridisation of resonant modes with opposite parity. This purely geometrical approach produces a pronounced linear chiroptical response, with an experimentally measured CD reaching 0.43. We further investigate nonlinear optical response through the study of third-harmonic generation (THG) under circularly polarized excitation. We find that nonlinear response is considerably stronger than its linear counterpart, exhibiting a pronounced preference for RCP excitation, with the generated THG signal enhanced by more than two orders of magnitude compared with an unpatterned silicon membrane. The resulting nonlinear CD reaches 0.91, demonstrating highly selective frequency conversion governed by the handedness of the incident light.

\begin{figure*}[!htbp]
    \centering
    \includegraphics[width=0.99\linewidth]{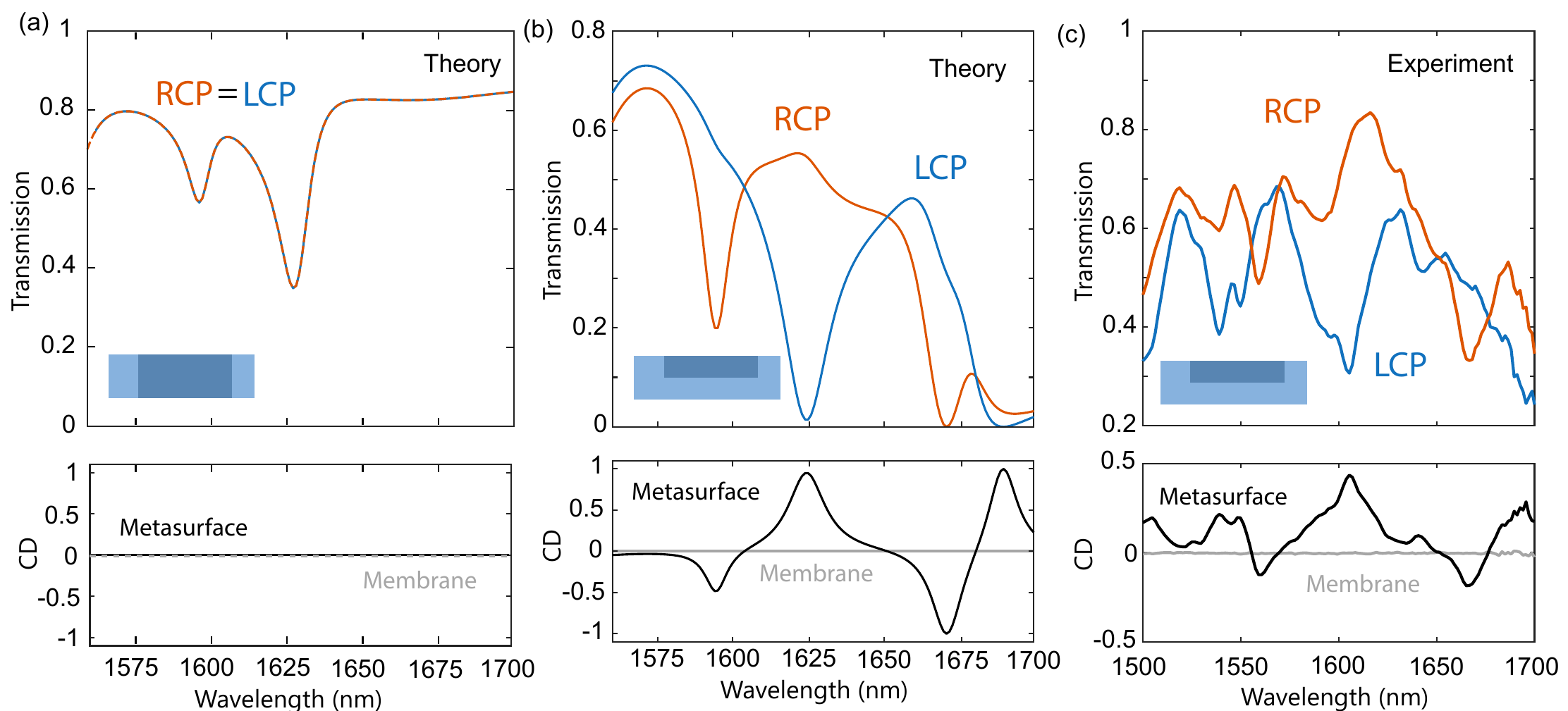}
    \caption{ \textbf{Linear transmission.} Simulated transmission of LCP and RCP light for a metasurface (a) without and (b) with a remaining Si layer beneath the patterned membranes. (c) Experimental transmission of LCP and RCP light for the metasurface with a Si layer beneath the membranes. The bottom panels show the linear CD of each metasurface (black curves). The grey curve in the panels shows the CD of the unpatterned membrane.}
    \label{fig:linear}
\end{figure*}

Our results establish partially etched, suspended membrane metasurfaces as a simple and versatile platform for chiral metaphotonics, providing a route to strong linear and nonlinear chiroptical responses without the need for additional material layers or complex three-dimensional meta-atom fabrication. This approach opens opportunities for handedness-selective frequency conversion, compact chiral light sources, and nonlinear polarization control. Furthermore, the ability to engineer out-of-plane symmetry directly through the membrane geometry may provide a useful platform for exploring more complex phenomena in topological~\cite{yin2023topological} and non-Hermitian photonics~\cite{kang2025janus}.

Figure~\ref{fig:concept}a illustrates the metasurface design, consisting of a free-standing Si membrane of thickness $h$ patterned with a square lattice of $C_4$-symmetric, four-lobed apertures. The apertures are described by the contour $r(\phi)=R[1+\delta\cos{4(\phi+\theta)}]$ and etched to a depth $d$, with a lattice period $\Delta$. Figure~\ref{fig:concept}b shows a scanning electron microscopy (SEM) image of the fabricated metasurface, confirming the realization of the designed aperture geometry. For $\theta\neq m\pi/4$ ($m$ an integer), the apertures are rotated relative to the lattice axes, breaking the in-plane mirror symmetries and resulting in a two-dimensionally chiral pattern.  For a fully etched membrane ($d=h$), the out-of-plane mirror symmetry is preserved, enforcing equal co-polarised transmission coefficients for LCP and RCP excitation and, consequently, zero linear CD. Here, this symmetry is broken intrinsically by partially etching the apertures, leaving a Si layer of thickness $h-d$ beneath them. 
The structure becomes three-dimentionally chiral. 
This reduces the point-group symmetry from $C_{4h}$ to $C_4$ and enables the hybridisation of resonant modes with opposite parity. Further details of the metasurface fabrication process are provided in Section S1 of the Supporting Information. However, symmetry breaking alone is insufficient to produce a nonzero linear CD: dissipation is also required in a reciprocal $C_4$ metasurface, with fabrication-related losses, including surface roughness, structural disorder and scattering, providing the necessary dissipation~\cite{Kumar2025chiral,Kumar2026}. We define the linear
circular dichroism as
\begin{equation}
    \mathrm{CD}^{\omega}=\frac{T_{\mathrm{R}}-T_{\mathrm{L}}}{T_{\mathrm{R}}+T_{\mathrm{L}}},
    \label{eq:CD:lin}
\end{equation}
where $T_{R(L)}$ is the transmission of RCP (LCP) light. For the nonlinear response, third-harmonic circular dichroism $\mathrm{CD}^{3\omega}$ is defined as
\begin{equation}
    \mathrm{CD}^{3\omega} = \frac{I_{\mathrm{R}}^{3\omega} - I_{\mathrm{L}}^{3\omega}}{I_{\mathrm{R}}^{3\omega} + I_{\mathrm{L}}^{3\omega}}
    \label{eq:CD:nonlin}
\end{equation}
where $I^{3\omega}_{\mathrm{R(L)}}$ is the intensity of the third-harmonic signal generated under RCP (LCP) pumping.

\begin{figure*}[!htbp]
    \centering
    \includegraphics[width=0.99\linewidth]{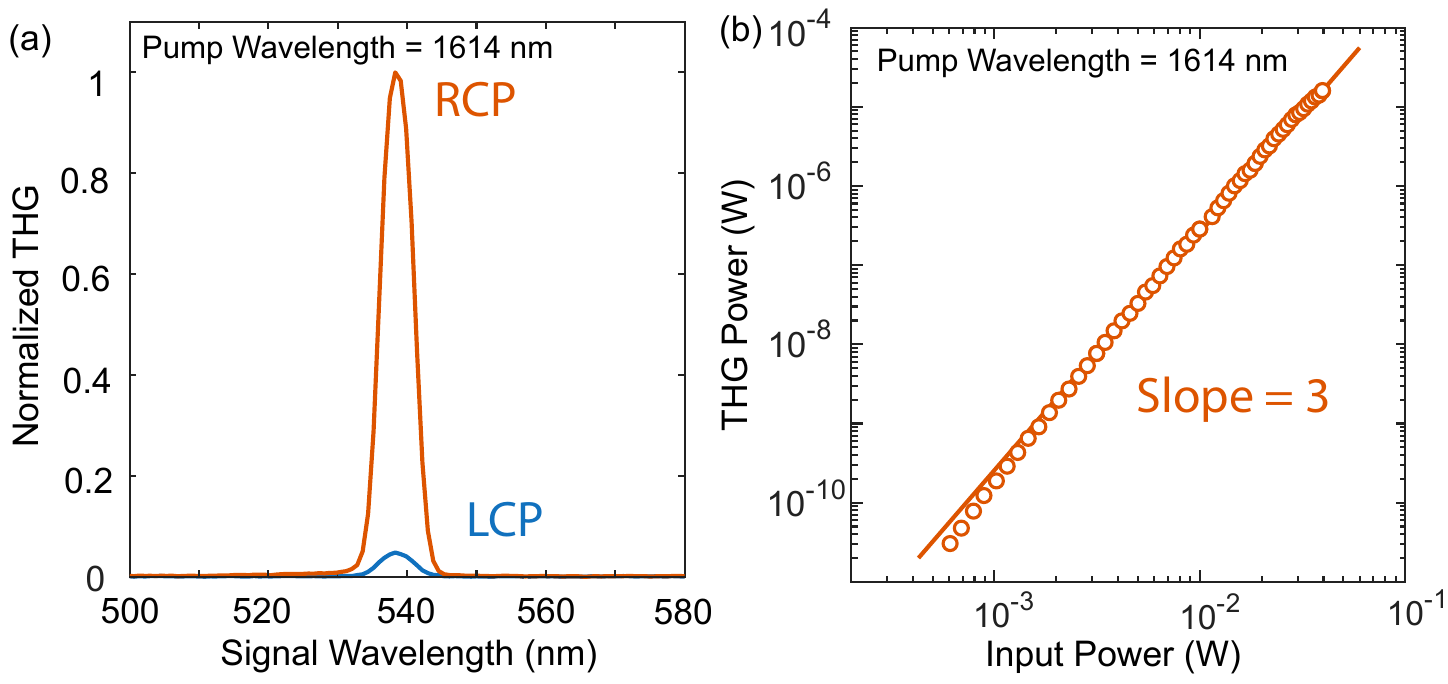}
    \caption{\textbf{Experimental third-harmonic generation}. (a) Third-harmonic spectra from the unperturbed metasurface pumped at a wavelength of 1614 nm with RCP (red) and LCP (blue) excitation. (b) Third-harmonic power as a function of pump power for the unperturbed metasurface under RCP excitation, with excitation wavelength at 1614 nm.}
    \label{fig:nonlinear_spec}
\end{figure*}

We first investigate the linear optical response of the suspended membrane metasurface in the absence of the underlying silicon layer. All numerical simulations were performed using the geometrical parameters of the metasurface described above. Details of the numerical methods, material parameters, and simulation settings are provided in the Supporting Information (Section S2). Figure~\ref{fig:linear}(a) shows the calculated transmission spectra for right- and left-circularly polarized (RCP and LCP) incident light. Two pronounced resonances are observed within the investigated wavelength range. Despite these resonant features, the transmission spectra for the two polarizations are identical, indicating that the metasurface exhibits no CD. This behaviour originates from the out-of-plane mirror symmetry of the suspended membrane, which ensures equivalent optical responses under RCP and LCP excitation. Accordingly, the linear CD, calculated using Eq.~\ref{eq:CD:lin}, is exactly zero across the entire wavelength range for both the patterned metasurface and the unpatterned membrane, as shown in the bottom panel of Fig.~\ref{fig:linear}(a). These results demonstrate that the resonant response of the patterned membrane alone is insufficient to produce a chiral optical response when the out-of-plane mirror symmetry is preserved.

We next investigate how breaking this symmetry affects the linear optical response by introducing an additional silicon layer beneath the patterned membrane. The corresponding transmission spectra for RCP and LCP excitation are presented in Fig.~\ref{fig:linear}(b). In contrast to the suspended configuration, the presence of the underlying silicon layer breaks the out-of-plane mirror symmetry and modifies the electromagnetic environment of the metasurface. Consequently, the resonant transmission features shift relative to those of the suspended membrane, and their spectral profiles become strongly dependent on the incident circular polarization. The resulting difference between the RCP and LCP transmission spectra gives rise to a pronounced chiroptical response. This behaviour is further illustrated in the bottom panel of Fig.~\ref{fig:linear}(b), which shows the calculated linear CD. The CD exhibits a strong wavelength dependence, reaching values close to the limiting values of (-1) and (+1) within the investigated spectral range. Thus, introducing a single silicon layer beneath an otherwise achiral membrane provides a simple means of inducing a strong linear chiral response without modifying the in-plane geometry of the metasurface.

To validate these theoretical predictions, we experimentally characterise fabricated metasurfaces based on the investigated design. The sample is illuminated with RCP and LCP light, and the corresponding transmitted signals are measured. Further details of the experimental setup and measurement procedure are provided in the Supporting Information Section S3. Figure~\ref{fig:linear}(c) presents the measured transmission spectra for the two circular polarizations. For comparison, we also measure the transmission through an unpatterned silicon membrane, for which the linear CD is zero, as shown in the bottom panel of Fig.~\ref{fig:linear}(c). This control measurement confirms that the experimental setup does not introduce a measurable spurious circular dichroism, supporting the reliability of the observed polarization-dependent response of the patterned metasurface.

Consistent with the numerical simulations, the experimental spectra exhibit pronounced polarization-dependent differences in the vicinity of the resonances, confirming the emergence of a chiral optical response in the presence of the underlying silicon layer. The overall spectral behaviour agrees qualitatively with the theoretical predictions, although the resonant features and the contrast between the two polarizations are less pronounced in the experimental data.

\begin{figure*}[!htbp]
    \centering
    \includegraphics[width=0.99\linewidth]{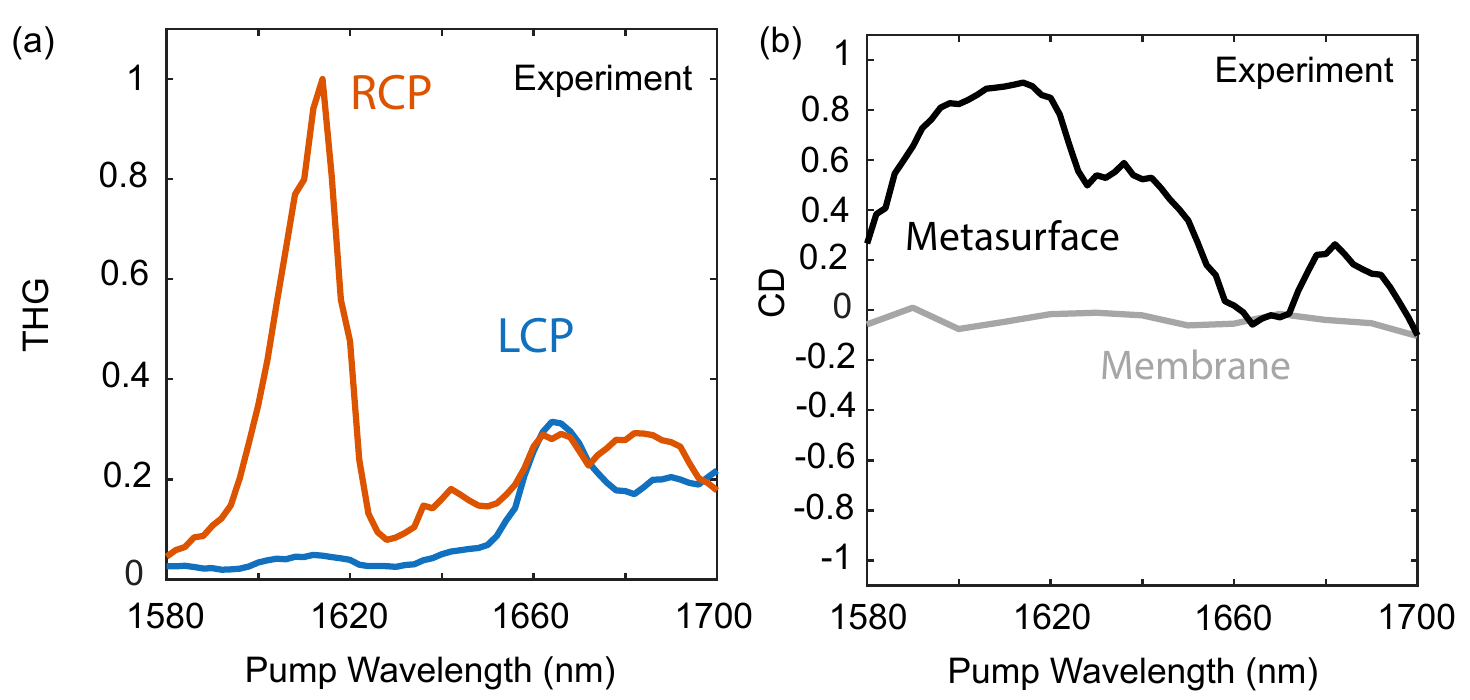}
    \caption{\textbf{Experimental chiral nonlinear response}. (a) Measured third harmonic signal from the metasurface under LCP (blue) and RCP (red) excitation. (b) Measured nonlinear CD from the metasurface and unpatterned membrane.}
    \label{fig:nonlinear_CD:exp}
\end{figure*}

The experimentally determined linear CD, calculated using Eq.~\ref{eq:CD:lin}, is shown in the bottom panel of Fig.~\ref{fig:linear}(c). The measured CD exhibits a spectral dependence similar to that obtained in the simulations, confirming that the symmetry-breaking mechanism is robust under experimental conditions. However, its magnitude is reduced compared with the idealised theoretical predictions, with the measured CD ranging from approximately (-0.18) to (0.43). We attribute this quantitative discrepancy primarily to fabrication imperfections, which can modify the geometry of the patterned features and affect the resonant optical response. Nevertheless, the observation of a substantial linear CD experimentally demonstrates that breaking the out-of-plane mirror symmetry of the membrane metasurface is an effective approach to achieving a strong chiral response in an otherwise achiral dielectric structure.

Having established the emergence of a strong chiral response in the linear regime, we next investigate the nonlinear optical properties of the metasurface, focusing on THG. In these experiments, the metasurface is excited from the silicon-layer side using a tunable femtosecond laser in the NIR spectral range, with both RCP and LCP excitation. The generated third-harmonic signal, corresponding to frequency tripling of the incident light, is collected in the visible spectral range. Further details of the experimental setup and measurement procedure are provided in the Supporting Information Section S3. 

Figure~\ref{fig:nonlinear_spec}(a) shows the measured third-harmonic spectra under RCP and LCP excitation at a pump wavelength of 1614 nm. A pronounced difference between the two polarizations is observed, with RCP excitation producing a substantially stronger third-harmonic signal than LCP excitation. This pronounced polarization dependence demonstrates that the chiral response of the metasurface extends beyond linear transmission to nonlinear frequency conversion. Notably, the strong THG signal under RCP excitation indicates that the metasurface supports a highly selective nonlinear response to the handedness of the incident light.

To further verify the nonlinear origin of the observed signal, we measure the dependence of the generated third-harmonic power on the incident pump power under RCP excitation. As shown in Fig.~\ref{fig:nonlinear_spec}(b), the THG power exhibits a cubic dependence on the pump power, consistent with the expected behaviour of a third-order nonlinear optical process. This cubic scaling confirms that the measured signal originates from third-harmonic generation. At an incident pump power of 39.6 mW, the maximum measured absolute THG conversion efficiency reaches (4 $\cdot$ 10$^{-4}$).

To further investigate the influence of resonant excitation on the nonlinear chiral response, we examine the wavelength dependence of THG in the membrane metasurface. The sample is excited from the silicon-layer side using a tunable femtosecond laser over a pump wavelength range of 1580–1700 nm, with both LCP and RCP excitation. At each pump wavelength, the corresponding THG spectrum is recorded, and its maximum intensity is extracted to characterise the wavelength-dependent nonlinear response. Further details of the experimental setup and measurement procedure are provided in the Supporting Information.

\begin{figure*}[!htbp]
    \centering
    \includegraphics[width=0.99\linewidth]{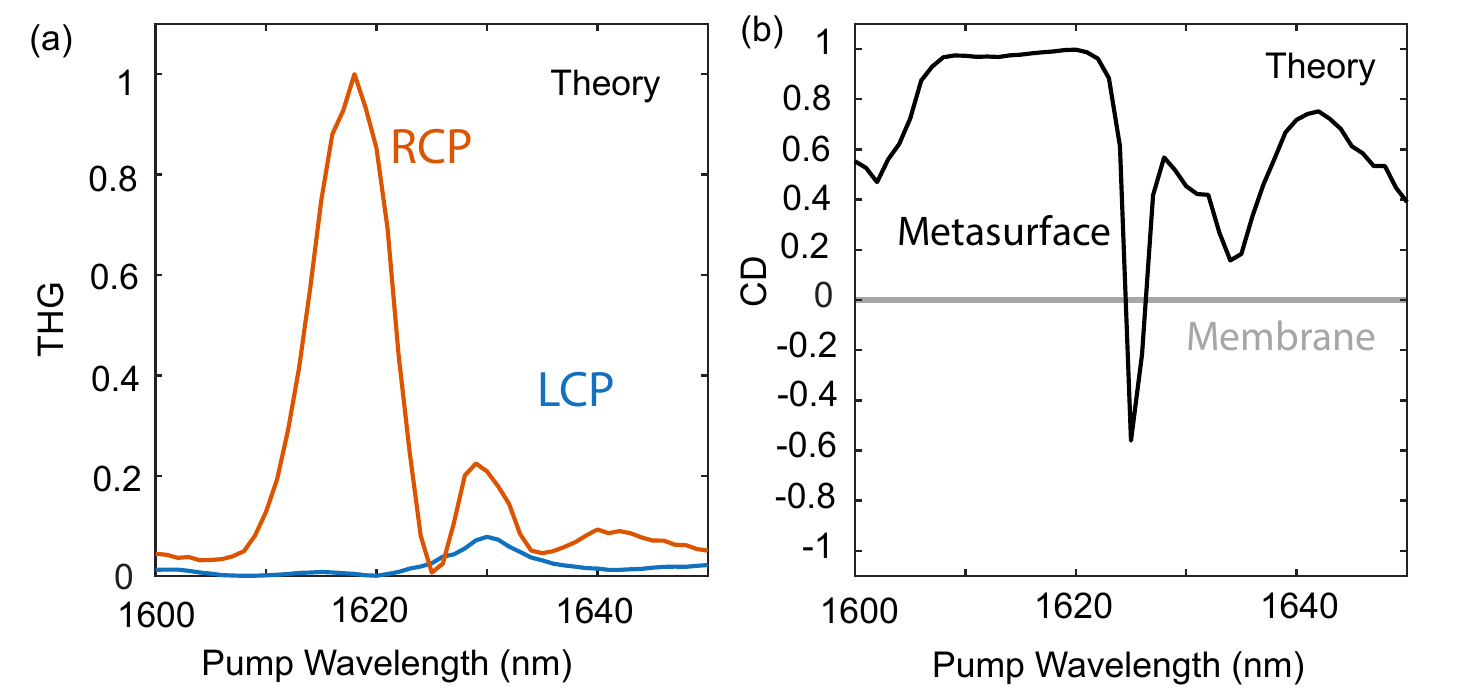}
    \caption{\textbf{Theoretical chiral nonlinear response}. (a) Calculated third harmonic signal from the metasurface under LCP (blue) and RCP (red) excitation. (b) Calculated nonlinear CD from the metasurface and unpatterned membrane.}
    \label{fig:nonlinear_CD:theory}
\end{figure*}

Figure~\ref{fig:nonlinear_CD:exp}(a) shows the maximum THG signal as a function of pump wavelength for both circular polarizations. The measured response exhibits a pronounced dependence on the handedness of the incident light, with RCP excitation producing a consistently stronger THG signal across the investigated spectral range. In particular, the RCP-pumped THG exhibits a pronounced enhancement around 1614 nm, consistent with the strong nonlinear response observed at a pump wavelength of 1614 nm in Fig.~\ref{fig:nonlinear_spec}a. In contrast, the THG signal under LCP excitation remains considerably weaker and exhibits only a relatively small enhancement around 1665 nm. These results demonstrate that the nonlinear response is strongly dependent on both the pump wavelength and the polarization handedness, with the strongest THG enhancement occurring for RCP excitation near 1614 nm.

To further verify that the observed polarization-dependent response originates from the patterned metasurface rather than the underlying membrane, we perform the same wavelength-dependent measurements on an unpatterned silicon membrane. In this control sample, the THG signals generated under LCP and RCP excitation exhibit no pronounced resonant features or appreciable differences in their spectral dependence (see Supporting Information Section S4). This behaviour confirms that the pronounced polarization selectivity observed in the patterned structure is associated with the metasurface. In addition to the pronounced polarization selectivity, the patterned metasurface exhibits a substantial enhancement of THG compared with the unpatterned silicon membrane. The maximum THG signal is enhanced by two orders of magnitude under RCP excitation, highlighting the role of the resonant metasurface in enhancing nonlinear frequency conversion. 

To quantify the nonlinear chiral response, we calculate the nonlinear CD using Eq.~\ref{eq:CD:nonlin}. The resulting experimental spectra for the patterned metasurface and the unpatterned membrane are presented in Fig.~\ref{fig:nonlinear_CD:exp}(b). The unpatterned membrane exhibits a nonlinear CD close to zero throughout the investigated wavelength range, consistent with the absence of a measurable polarization-dependent THG response. In contrast, the metasurface exhibits a pronounced wavelength-dependent nonlinear CD, reaching a maximum value of 0.91 near the RCP-enhanced THG resonance. This large value indicates a strong preference for THG under RCP excitation and demonstrates that the metasurface supports a highly selective nonlinear chiral response. Importantly, the close-to-zero nonlinear CD of the unpatterned membrane confirms that this effect arises from the patterned metasurface rather than an intrinsic polarization-dependent response of the membrane.

To further elucidate the experimentally observed nonlinear chiroptical response, we perform numerical simulations of the THG response for both the patterned metasurface and the unpatterned membrane, using the geometrical parameters of the fabricated structures. Details of the numerical modelling are provided in the Supporting Information. As shown in Fig.~\ref{fig:nonlinear_CD:theory}(a), the THG response of the patterned metasurface exhibits pronounced resonant enhancement, with RCP excitation producing a substantially stronger signal around the corresponding resonance. In contrast, LCP excitation gives rise to a weaker enhancement associated with the other resonant mode. This behaviour is in qualitative agreement with the experimental observations, although some differences in the spectral response are apparent, which may arise from fabrication imperfections and deviations of the experimental structures from the idealised geometries used in the simulations. We further quantify the simulated chiroptical response by calculating the nonlinear CD using Eq.~\ref{eq:CD:nonlin}, as shown in Fig.~\ref{fig:nonlinear_CD:theory}(b). The nonlinear CD is positive over most of the spectral range, indicating the dominance of RCP-driven THG, with a narrow spectral region exhibiting the opposite response. Notably, the nonlinear CD approaches unity near the resonant enhancement of the patterned metasurface, demonstrating nearly complete discrimination between the two excitation handednesses. In contrast, the unpatterned membrane exhibits zero nonlinear CD, consistent with the absence of two-dimensional chirality, as its geometry and its crystal axes, aligned with the lattice, have in-plane mirror lines.

We have demonstrated that a suspended, substrate-free membrane metasurface can be transformed from achiral to strongly chiral by introducing a thin silicon layer beneath the patterned membrane, remaining after incomplete etching. This remaining layer breaks the out-of-plane mirror symmetry, and it enables the hybridisation of resonant modes with opposite parity, giving rise to pronounced chiroptical responses in both linear and nonlinear regimes. Experimentally, we have achieved linear CD of up to 0.43 and nonlinear CD of 0.91 in third-harmonic generation. These results highlight the effectiveness of symmetry breaking in controlling light–matter interactions and demonstrate a versatile approach to enhancing handedness-selective nonlinear frequency conversion. Our findings provide a promising route towards compact nonlinear photonic devices for the generation and manipulation of chiral light.

\section*{Acknowledgements}

Y.K. acknowledges a support from the Australian Research Council (Grant DP210101292) and the International Technology Center Indo-Pacific (ITC IPAC) via Army Research Office (contract FA520923C0023). J.R. acknowledges support from the National Research Foundation (NRF) grant (RS-2024-00356928) funded by the Ministry of Science and ICT (MSIT) of the Korean government. This work used the ACT node of the NCRIS-enabled Australian National Fabrication Facility (ANFF-ACT).


\newpage

\bibliography{ref}

\end{document}